\documentclass[aps,twocolumn,showpacs,preprintnumbers,amsmath,amssymb,superscriptaddress,floatfix,nofootinbib]{revtex4}
\usepackage{graphicx}
\usepackage{epsfig}
\usepackage{epstopdf}
\usepackage{hyperref}
\usepackage{amsmath}
\usepackage{amsfonts}
\usepackage{amssymb}

\usepackage{graphicx,color,dcolumn,booktabs}
\usepackage{txfonts}
\usepackage{amssymb}
\usepackage{indentfirst}
\usepackage{subfig}
\usepackage{float}
\usepackage{multirow}
\usepackage{color}

\allowdisplaybreaks
\begin{document}

\title{Masses of doubly charmed baryons in the extended on-mass-shell renormalization scheme}

%\email{}

\author{Zhi-Feng Sun}
\affiliation{Departamento de
F\'{\i}sica Te\'orica and IFIC, Centro Mixto Universidad de
Valencia-CSIC Institutos de Investigaci\'on de Paterna, Aptdo.
22085, 46071 Valencia, Spain}

\author{M. J. Vicente Vacas}
\affiliation{Departamento de
F\'{\i}sica Te\'orica and IFIC, Centro Mixto Universidad de
Valencia-CSIC Institutos de Investigaci\'on de Paterna, Aptdo.
22085, 46071 Valencia, Spain}

\date{\today}

\begin{abstract}
In this work, we investigate the mass corrections of the doubly charmed baryons up to $N^2LO$ in the extended-on-mass-shell (EOMS) renormalization scheme, comparing with the results of heavy baryon chiral perturbation theory. 
We find that the terms from the heavy baryon approach are a subset of those obtained in the EOMS scheme. By fitting the lattice data, we can determine the parameters $\tilde{m}$, $\alpha$, $c_1$ and $c_7$ from the Lagrangian, while in the heavy baryon approach no information on $c_1$ can be obtained from the baryons mass.
 Correspondingly, the masses of $m_{\Xi_{cc}}$ and $m_{\Omega_{cc}}$ are predicted,  in the EOMS scheme, extrapolating the results from different values of the charm quark and the pion masses of the lattice QCD calculations.

\end{abstract}

\maketitle

\section{Introduction}
The doubly charmed baryons are composed of two charmed quarks and one light quark.
 The ones with quark components $ccu$, $ccd$ and $ccs$ are named as $\Xi_{cc}^{++}$, $\Xi_{cc}^{+}$ and $\Omega_{cc}^{++}$.
 Whereas most of their mass comes from their charm content, it is interesting to study the chiral corrections related to the light quark and their influence on the mass splitting.

In the past decades, there has been some experimental effort searching for the doubly charmed baryons \cite{Mattson:2002vu, Moinester:2002uw, Ocherashvili:2004hi, Engelfried:2005kd}, although the situation is still unsettled. $\Xi_{cc}^{+}(3520)$ was reported in $\Lambda_c^+K^-\pi^+$ channel by SELEX collaboration with the mass $3519\pm 1$ MeV and the width less than $5$ MeV. Later, this state was confirmed in $pD^+K^-$ channel by SELEX with a mass of $3518\pm 3$ MeV. SELEX also have the evidence of the $\Xi_{cc}^{++}$ baryons with masses of $3460$ MeV and $3780$ MeV which are detected in $\Lambda_c^+K^-\pi^+\pi^+$ mode. However, none of these states were confirmed by other experiments, such as FOCUS \cite{Ratti:2003ez}, BABAR \cite{Aubert:2006qw}, Belle \cite{Chistov:2006zj} and LHCb \cite{Aaij:2013voa}.

Theoretical studies of doubly charmed baryons have been performed with different approaches. Lattice QCD groups predict that the mass of $\Xi_{cc}$ is in the range $3.51\sim 3.67$ GeV, and that of $\Omega_{cc}^+$ in $3.68\sim 3.76$ GeV \cite{Lewis:2001iz, Na:2008hz, Liu:2009jc, Namekawa:2012mp, Alexandrou:2012xk, Padmanath:2015jea}. The quark model predictions of $\Xi_{cc}$ and $\Omega_{cc}$ masses are in the ranges of $3.48\sim 3.74$ GeV and $3.59\sim 3.86$ GeV, respectively \cite{Roncaglia:1995az, Ebert:1996ec, SilvestreBrac:1996wp, Tong:1999qs, Gerasyuta:1999pc, Itoh:2000um, Kiselev:2001fw, Narodetskii:2002ib, Ebert:2002ig, Vijande:2004at, Migura:2006ep, Albertus:2006ya, Roberts:2007ni,Karliner:2014gca}. In Ref. \cite{Brodsky:2011zs}, the mass splitting of doubly charmed baryons is studied in chiral perturbation theory (ChPT) considering heavy diquark symmetry.
In Ref. \cite{Sun:2014aya}, within the framework of heavy baryon ChPT, the mass corrections of doubly charmed baryons were studied up to N$^3$LO. See Ref.~\cite{Cheng:2015iom} for a review of the current situation on both the theoretical and the experimental side.

Before this latter work, the light baryons' mass corrections had been abundantly studied in ChPT (see reviews \cite{Scherer:2002tk, Bernard:1995dp}); In Refs. \cite{Guo:2008ns, Jiang:2014ena}, the mass corrections of singly heavy baryons were investigated.

Quantum chromodynamics (QCD) is the theory which describes the strong interaction. In the high energy regime, perturbative QCD works very well due to asymptotic freedom, while in the low energy region perturbation theory fails to converge. This  low energy region can be studied  constructing an effective Lagrangian based on the  the QCD symmetries and the relevant degrees of freedom. The corresponding theory is ChPT. The Lagrangian is expressed in terms of hadronic fields and organized in the form of a chiral expansion, i.e., an expansion in powers of momentum and light quark masses. When investigating the high order corrections, the chiral power counting scheme, proposed by Weinberg $et$ $al$ \cite{Weinberg:1991um, Ecker:1994gg} is used. However, this leads to some difficulties in the baryon sector. Namely, the loop diagrams  violate the power counting due to the non-vanishing baryon masses in chiral limit. To solve this issue, various schemes have been proposed, such as heavy baryon chiral perturbation theory \cite{Jenkins:1990jv}, infrared baryon chiral perturbation theory \cite{Becher:1999he, Kubis:2000zd} and the extended-on-mass-shell (EOMS) approach \cite{Fuchs:2003qc}. See Ref.~ \cite{Geng:2013xn}, for a brief explanation and comparison of the three renormalization schemes. 
The heavy baryon approach was motivated by the methods used in heavy quark effective field theory, in which the baryon is treated to be extremely heavy and acts as a static source. Henceforth, one can take the non relativistic limit and make the expansion in powers of the inverse baryon mass. Within the infrared baryon ChPT, it is used that the infrared singular part of the loop integral conserves the Weinberg's power counting rule. In the EOMS scheme, after calculating the loops covariantly, the power counting breaking terms are subtracted which we will discuss below.

In this work, we will use ChPT with the EOMS renormalization scheme, which has been particularly successful for the light baryon masses~\cite{ MartinCamalich:2010fp}, to investigate the masses of doubly charmed baryons, and make a comparison with the heavy baryon ChPT results of Ref. \cite{Sun:2014aya}.

Our work is organized as follows. In Sec. II, the chiral Lagrangian is introduced. We will calculate the doubly charmed baryon masses in the EOMS scheme comparing with the expression from the heavy baryon approach in Sec. III and Sec. IV. Then, we show the numerical results in Sec. V. Finally, a short summary is given.

\section{the effective Lagrangian}
In Ref. \cite{Sun:2014aya}, the effective Lagrangian describing the interaction of doubly charmed baryons and the Goldstone bosons was constructed. The relevant pieces are
\begin{eqnarray}
\mathcal{L}^{(1)}&=&\bar{\psi}_0(i{D\!\!\!\slash}-m_0+\frac{g_A}{2}\gamma^\mu\gamma_5u_\mu)\psi_0,\label{l1}\label{eq:L1}\\
\nonumber 
\mathcal{L}^{(2)}&=&c_1\bar{\psi}_0\langle
\chi_+\rangle\psi_0-\left\{\frac{c_2}{8m_0^2} \bar{\psi}_0\langle u_\mu
u_\nu \rangle \{ D^\mu,D^\nu\}\psi_0+h.c.\right\}\\\nonumber
&&-\left\{\frac{c_3}{8m_0^2} \bar{\psi}_0\{ u_\mu, u_\nu\}  \{
D^\mu,D^\nu\}\psi_0+h.c.\right\}+\frac{c_4}{2}\bar{\psi}_0\langle
u^2\rangle\psi_0\\\nonumber &&+\frac{c_5}{2}\bar{\psi}_0 u^2\psi_0+
\frac{ic_6}{4}\bar{\psi}_0\sigma^{\mu\nu}[u_\mu,u_\nu]\psi_0+c_7\bar{\psi}_0\hat{\chi_+}\psi_0\\
&&+\frac{c_8}{8m_0}\bar{\psi}_0\sigma^{\mu\nu}f^+_{\mu\nu}\psi_0+\frac{c_9}{8m_0}\bar{\psi}_0\sigma^{\mu\nu}\langle
f^+_{\mu\nu}\rangle\psi_0\label{eq:L2}.
\end{eqnarray}
Note that the fields here are the bare fields and the mass $m_0$ is the bare mass of the considered doubly heavy baryons. $U$ and $u$ which incorporate the pseudoscalar meson field are defined as
\begin{eqnarray}
U=u^2=\exp\left(i\frac{\phi_0(x)}{F_0}\right),
\end{eqnarray}
where $\phi(x)$ is expressed as
\begin{eqnarray}
\phi_0(x)=\left(
          \begin{array}{ccc}
            \pi_0^0+\frac{1}{\sqrt{3}}\eta_0 & \sqrt{2}\pi_0^+ & \sqrt{2}K_0^+ \\
            \sqrt{2}\pi_0^- & -\pi_0^0+\frac{1}{\sqrt{3}}\eta_0 & \sqrt{2}K_0^0 \\
            \sqrt{2}K_0^- & \sqrt{2}\bar{K_0^0} & -\frac{2}{\sqrt{3}}\eta_0 \\
          \end{array}
        \right).
\end{eqnarray}
The doubly heavy baryon field $\psi_0$ with spin $\frac{1}{2}$ is a
column vector in the flavor space, i.e.
\begin{eqnarray}
\psi_0 &=&\left(
          \begin{array}{c}
            \Xi_{cc}^{++} \\
            \Xi_{cc}^{+} \\
            \Omega_{cc}^+ \\
          \end{array}
        \right).
\end{eqnarray}
In Eq. \eqref{eq:L1}-\eqref{eq:L2}, $\chi$, $\chi_\pm$, $f_{\mu\nu}^R$, $f_{\mu\nu}^L$ $f_{\mu\nu}^\pm$, $u_\mu$, $\Gamma_\mu$, $D_\mu$ have the following definition
\begin{eqnarray}
\chi&=&2B_0(s+ip),\\
\chi_{\pm}&=&u^\dag\chi u^\dag\pm u\chi^\dag u,\\
f^R_{\mu\nu}&=&\partial_\mu r_\nu-\partial_\nu r_\mu-i[r_\mu,r_\nu],\\
f^L_{\mu\nu}&=&\partial_\mu l_\nu-\partial_\nu l_\mu-i[l_\mu,l_\nu],\\
f^\pm_{\mu\nu}&=&u^\dag f^R_{\mu\nu} u\pm uf^L_{\mu\nu} u^\dag,\\
u_\mu&=&i[u^\dag(\partial_\mu-ir_\mu)u-u(\partial_\mu-il_u)u^\dag],\\
\Gamma_\mu&=&\frac{1}{2}[u^\dag(\partial_\mu-ir_\mu)u+u(\partial_\mu-il_u)u^\dag],\\
D_\mu&=&\partial_\mu+\Gamma_\mu-iv^{(s)}_\mu,
%\\D^\prime_\mu&=&\partial_\mu-\Gamma_\mu+iv^{(s)}_\mu,
\end{eqnarray}
where $r_\mu=v_\mu+a_\mu$, $l_\mu=v_\mu-a_\mu$, and $v_\mu,
v^{(s)}_\mu, a_\mu, s, p$ are external $c$-number fields.

By introducing the renormalized fields through
\begin{eqnarray}
&&\psi=\frac{\psi_0}{\sqrt{Z_{\psi}}},\, \pi^{\pm, 0}=\frac{\pi^{\pm, 0}_0}{\sqrt{Z_{\pi}}},\, K^{\pm, 0}=\frac{K^{\pm, 0}_0}{\sqrt{Z_{K}}},\nonumber\\
&&\bar{K}^0=\frac{\bar{K}^0_0}{\sqrt{Z_{K}}},\, \eta=\frac{\eta_0}{\sqrt{Z_{\eta}}},
\end{eqnarray}
the Lagrangian of bare fields could be expressed as the sum of basic and counterterm Lagrangians 
\begin{eqnarray}
\mathcal{L}=\mathcal{L}_{basic}+\mathcal{L}_{counterterm},
\end{eqnarray}
where 
\begin{eqnarray}
\mathcal{L}_{basic}&=&\bar{\psi}_a(i{\partial\!\!\!\slash}-m)\psi_a-\bar{\psi}_a\frac{g_A}{2}\gamma_\mu\gamma_5\partial^\mu\phi_{ab} \psi_b+\cdots,\nonumber\\
\mathcal{L}_{counterterm}&=&(Z_a-1)\bar{\psi}_ai{\partial\!\!\!\slash}\psi_a-(Z_a-1)\bar{\psi}_am\psi_a\nonumber\\
&&-Z_a\bar{\psi}_a\delta m_a\psi_a+\cdots.
\end{eqnarray}
Here, $\phi_{ab}$ is the renormalized field.
The sum is performed for the repeated indices.
$a$ and $b$ are the indices in the flavor space ($a,b=1,2,3$ denoting $\Xi^{++}_{cc}$, $\Xi^{+}_{cc}$, $\Omega^{++}_{cc}$ respectively), $m$ is the mass in the chiral limit, and $Z_a$ is the wave function renormalization constant. Note that here we only show the expression of the lowest order.

%\section{Power counting}
%According to Weinberg's power counting rule, the order of a diagram is expressed as 
%\begin{eqnarray}
%D=4 N_L-2I_\pi-I_N+\sum_{k=1}^{\infty}kN_{k},\label{eq:power_counting}
%\end{eqnarray}
%where $N_L$, $I_\pi$, $I_N$ and $N_k$ denote the number of independent loop momenta, internal meson lines, internal baryon lines and vertices of the Lagrangian of order $k$.

%When calculating the meson baryon loop integrals, we find that the result breaks the power counting of Eq. \eqref{eq:power_counting}. Extend-on-mass-shell renormalization scheme is to deal with this kind of problem, in which the power counting breaking terms is subtracted by introducing suitable counter terms. In this work, we will use this scheme comparing with the heavy baryon approach.

\begin{figure}[htpb]
\centering
\includegraphics[width=0.9\linewidth]{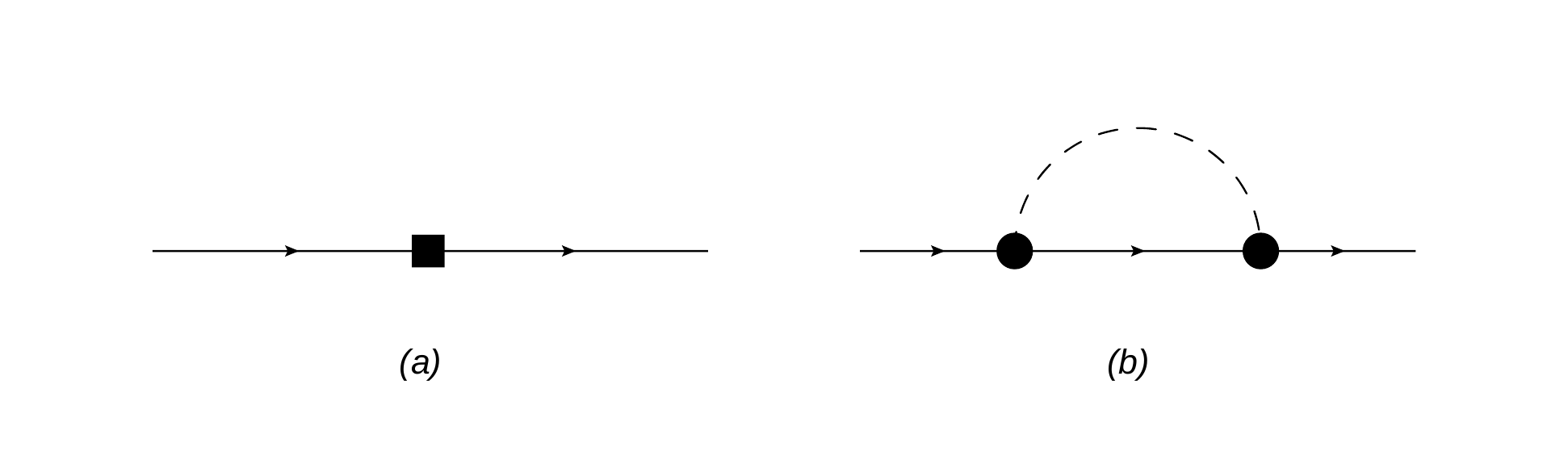}
\caption{The Feynman diagrams which contribute to the self-energy of
doubly charmed baryon. The solid and dashed lines denote the doubly
charmed baryons and Goldstone bosons. The solid dot
and black box denote the vertices from the ${\cal O}(p^1, p^2)$
Lagrangians respectively. \label{fig:fynmandiagram}}
\end{figure}

\section{the doubly charmed baryons masses}
The full propagator has the form of
\begin{eqnarray}
S(p)&=&\frac{1}{p\!\!\!\slash-m_0-\Sigma(p\!\!\!\slash)}\nonumber\\
&=&\frac{1}{p\!\!\!\slash-m-\Sigma_r(p\!\!\!\slash)}\nonumber\\
&\Rightarrow &\frac{1}{p\!\!\!\slash-m-\Sigma_r(p\!\!\!\slash)\big|_{p\!\!\!\slash=m_a}-(p\!\!\!\slash-m_a)[\Sigma_r(p\!\!\!\slash)]^\prime\big|_{p\!\!\!\slash=m_a}}\nonumber\\
&=&\frac{1}{1-[\Sigma_r(p\!\!\!\slash)]^\prime\big|_{p\!\!\!\slash=m_a}}\frac{1}{p\!\!\!\slash-m_a}\nonumber\\
&=&\frac{Z_a}{p\!\!\!\slash-m_a},
\end{eqnarray}
where $\Rightarrow$ means the case in the limit of $p\!\!\!\slash \to m_a$.
In the above function, $Z_a$ is the wave function renormalization constant which is defined as the residue at $p\!\!\!\slash\to m_a$, $\Sigma_r(p\!\!\!\slash)$ corresponds to the baryon self-energy, and $m_a=m+\Sigma_r(p\!\!\!\slash)\big|_{p\!\!\!\slash=m_a}$ is the physical mass of the baryon. The contributions related to FIG. \ref{fig:fynmandiagram} are 
\begin{eqnarray}
\Sigma^{(1)}_{a}&=&-2c_1(2m_K^2+m_\pi^2)-2c_7[\chi_{aa}-\frac{1}{3}(2m_K^2+m_\pi^2)],\\
\Sigma^{(2)}_{\lambda,ab}&=&iC^\lambda_{ab}\frac{g_A^2}{4F_\lambda^2}\int \frac{d^nk}{(2\pi)^n} \frac{k\!\!\!\slash \gamma_5(p\!\!\!\slash-k\!\!\!\slash+m)k\!\!\!\slash\gamma_5}{(p-k)^2-m^2+i\epsilon}\frac{1}{k^2-M_\lambda^2+i\epsilon}\nonumber\\
&=&-C^\lambda_{ab}\frac{g_A^2}{4F_\lambda^2}\left\{-(p^2-m^2)p\!\!\!\slash\frac{1}{2p^2}[(p^2-m^2+M_\lambda^2)\right.\nonumber\\
&&\times I_{\lambda a}(-p,0)+I_a(0)-I_\lambda(0)]+(p\!\!\!\slash+m)M_\lambda^2\nonumber\\
&&\times I_{\lambda a}(-p,0)+(p\!\!\!\slash+m)I_a(0)\large\}.\label{eq:sigma2}
\end{eqnarray}
The index $\lambda=\pi,K,\eta$. $M_\lambda$ and $F_\lambda$ are mass and decay constant of the meson $\lambda$, respectively.
The coefficient $C_{ab}^\lambda$ is the element of the matrices
\begin{eqnarray}
C^\pi=\left(\begin{array}{ccc}
1 & 2 & 0 \\ 
2 & 1 & 0 \\ 
0 & 0 & 0
\end{array} \right),\,\,
C^K=\left(\begin{array}{ccc}
0 & 0 & 2 \\ 
0 & 0 & 2 \\ 
2 & 2 & 0
\end{array} \right),\,\,
C^\eta=\left(\begin{array}{ccc}
\frac{1}{3} & 0 & 0 \\ 
0 & \frac{1}{3} & 0 \\ 
0 & 0 & \frac{4}{3}
\end{array} \right).
\end{eqnarray}
The form of the integral $I_{\lambda,a}$, $I_{a}$ and $I_{\lambda}$ are shown in the appendix.
Removing the infinite piece in the loop integral using $\widetilde{MS}$ scheme, we denote the corresponding finite part of the loop integral by $\Sigma^{(2)}_{r,\lambda,ab}$.
We extract the term breaking the power counting rule from Eq. \eqref{eq:sigma2} as follows
\begin{eqnarray}
\Sigma_r^{break}&=&C^\lambda_{ab}\frac{g_A^2}{32\pi^2F_\lambda^2}\left[-\frac{1}{4m}(p^2-m^2)^2+mM_\lambda^2\right].\label{breakpowercounting}
\end{eqnarray}
Thus, the mass of the doubly charmed baryons is expressed as
\begin{eqnarray}
m_a&=&m+\Sigma^{(1)}_{a}+\sum_{b=1}^{3}\sum_{\lambda=\pi,K,\eta}\Sigma^{(2)}_{r,\lambda,ab}|_{p\!\!\!\slash\to m_a}+\delta m_a\nonumber\\
&\doteq&m-2c_1(2m_K^2+m_\pi^2)-2c_7\left[\chi_{aa}-\frac{1}{3}(2m_K^2+m_\pi^2)\right]\nonumber\\
&&+\sum_{b=1}^{3}\sum_{\lambda=\pi,K,\eta}(-)C^\lambda_{ab}\frac{g_A^2}{4F_\lambda^2}2mM_\lambda^2\frac{1}{(4\pi)^2}\left[-1+\frac{M_\lambda^2}{2m^2}\ln\frac{M_\lambda^2}{m^2}\right.\nonumber\\
&&\left.+\frac{M_\lambda\sqrt{4m^2-M_\lambda^2}}{m^2}\arccos\frac{M_\lambda}{2m} \right]+\delta m_a,
\end{eqnarray}
where $\delta m_a$ should be the negative of Eq. \eqref{breakpowercounting} after substituting $p\!\!\!\slash=m_a$, i.e.,
\begin{eqnarray}
\delta m_a&=&-\sum_{b=1}^{3}\sum_{\lambda=\pi,K,\eta}C^\lambda_{ab}\frac{g_A^2}{32\pi^2F_\lambda^2}mM_\lambda^2.
\end{eqnarray}
And then, we get
\begin{eqnarray}
m_a&=&m-2c_1(2m_K^2+m_\pi^2)-2c_7\left[\chi_{aa}-\frac{1}{3}(2m_K^2+m_\pi^2)\right]\nonumber\\
&&+\sum_{b=1}^{3}\sum_{\lambda=\pi,K,\eta}(-)C^\lambda_{ab}\frac{g_A^2}{4F_\lambda^2}2mM_\lambda^2\frac{1}{(4\pi)^2}\left[\frac{M_\lambda^2}{2m^2}\ln\frac{M_\lambda^2}{m^2}\right.\nonumber\\
&&\left.+\frac{M_\lambda\sqrt{4m^2-M_\lambda^2}}{m^2}\arccos\frac{M_\lambda}{2m} \right].\label{eq:mass_eoms1}
\end{eqnarray}
Next, we perform the expansion in powers of the Goldstone boson mass
\begin{eqnarray}
m_a&=&m-2c_1(2m_K^2+m_\pi^2)-2c_7\left[\chi_{aa}-\frac{1}{3}(2m_K^2+m_\pi^2)\right]\nonumber\\
&&-\sum_{b=1}^{3}\sum_{\lambda=\pi,K,\eta}C^\lambda_{ab}\frac{g_A^2}{32\pi^2F_\lambda^2}m\left[\frac{\pi M_\lambda^3}{m}+\cdots\right].\label{eq:mass_eoms2}
\end{eqnarray}
Here the ellipsis denotes the contribution of orders higher than 3.

\section{Comparison of doubly heavy baryon mass in EOMS and heavy baryon schemes}

Besides the EOMS scheme, heavy baryon chiral perturbation theory (HBChPT) has also been used to study the doubly heavy baryon masses. In this section, we will give a comparison of the results under both schemes. First, we give a brief discussion of HBChPT.

Considering the baryon mass is extremely heavy, we have the picture that the baryon is surrounded by a cloud of light mesons. In this case, the four-momentum of the baryon $p^\mu$ can be separated into a large piece and a soft residual component 
\begin{eqnarray}
p^\mu=mv^\mu+l^\mu,
\end{eqnarray} 
where $v^\mu$ is the four-velocity, and $v\cdot l\ll m$. Using the projection operator $\mathcal{P}_v^{\pm}\equiv(1+v\!\!\!\slash)/2$, one can define the velocity-dependent fields
\begin{eqnarray}
H=e^{imv\cdot x}\mathcal{P}_v^+\psi, \, \, h=e^{imv\cdot x}\mathcal{P}_v^-\psi,
\end{eqnarray}
which are also called light and heavy components, respectively.
Henceforth, the baryon field $\psi$ is expressed as 
\begin{eqnarray}
\psi=e^{-imv\cdot x}(H+h).
\end{eqnarray}
If projecting the equation of motion onto the $\mathcal{P}_v^{+}$ and $\mathcal{P}_v^{-}$ parts, one has
\begin{eqnarray}
(iv\cdot D+\frac{g_A}{2}u\!\!\!\slash_\perp\gamma_5+\cdots)H+(iD\!\!\!\slash_\perp+\frac{g_A}{2}v\cdot u\gamma_5\nonumber\\+\cdots)h=0,\label{eq:eom1}\\
(iD\!\!\!\slash_\perp-\frac{g_A}{2}v\cdot u\gamma_5+\cdots)H+(-iv\cdot D-2m+\frac{g_A}{2}u\!\!\!\slash_\perp\gamma_5\nonumber\\+\cdots)h=0\label{eq:eom2}
\end{eqnarray}
with $A^\mu_\perp=A^\mu-v\cdot A v^\mu$ and the ellipsis means higher order contributions.
After solving the Eq. \eqref{eq:eom2} for $h$ and inserting the result into Eq. \eqref{eq:eom1}, we arrive at
\begin{eqnarray}
&&(iv\cdot D+\frac{g_A}{2}u\!\!\!\slash_\perp\gamma_5+\cdots)H+(iD\!\!\!\slash_\perp+\frac{g_A}{2}v\cdot u\gamma_5+\cdots)\nonumber\\
&&\times (2m+iv\cdot D-\frac{g_A}{2}u\!\!\!\slash_\perp\gamma_5+\cdots)^{-1}(iD\!\!\!\slash_\perp-\frac{g_A}{2}v\cdot u\gamma_5\nonumber\\
&&+\cdots)H=0,
\end{eqnarray}
which represents the equation of motion of the field $H$. Consequently, the corresponding Lagrangian is
\begin{eqnarray}
\mathcal{L}_H&=&\bar{H}(iv\cdot D+\frac{g_A}{2}u\!\!\!\slash_\perp\gamma_5+\cdots)H+\bar{H}(iD\!\!\!\slash_\perp+\frac{g_A}{2}v\cdot u\gamma_5\nonumber\\
&&+\cdots)(2m+iv\cdot D-\frac{g_A}{2}u\!\!\!\slash_\perp\gamma_5+\cdots)^{-1}(iD\!\!\!\slash_\perp\nonumber\\
&&-\frac{g_A}{2}v\cdot u\gamma_5+\cdots)H
\end{eqnarray}
By expanding the above equation in powers of $1/m$, we obtain
\begin{eqnarray}
\mathcal{L}_H^{(1)}&=&\bar{H}(iv\cdot D+g_AS_v\cdot u)H,\nonumber\\
\mathcal{L}_H^{(2)}&=&c_1\langle \chi_+\rangle+\frac{c_2}{2}\langle (v\cdot
u)^2\rangle+c_3(v\cdot u)^2+\frac{c_4}{2}\langle
u^2\rangle\nonumber\\
&&+\frac{c_5}{2}u^2+\frac{c_6}{2}[S_v^\mu, S_v^\nu][u_\mu,
u_\nu]+c_7\hat{\chi_+}\nonumber\\
&&-\frac{ic_8}{4m}[S_v^\mu,
S_v^\nu]f^+_{\mu\nu}-\frac{ic_9}{4m}[S_v^\mu, S_v^\nu]\langle
f^+_{\mu\nu}\rangle\nonumber\\
&&+\frac{2}{m}(S_v\cdot D)^2-\frac{ig_A}{2m}\{S_v\cdot D, v\cdot
u\}\nonumber\\
&&-\frac{g_A^2}{8m}(v\cdot u)^2+\cdots.
\end{eqnarray}
Using the heavy baryon Lagrangian, one obtains the doubly heavy baryon mass up to chiral order three: 
\begin{eqnarray}
m_{Ha}&=&m-2c_1(2m_K^2+m_\pi^2)-2c_7(\chi_{aa}-\frac{1}{3}(2m_K^2+m_\pi^2))\nonumber\\
&&-\sum_{b=1}^{3}\sum_{\lambda=\pi,K,\eta}C^\lambda_{ab}\frac{g_A^2}{(4\pi F_\lambda)^2}\frac{\pi}{2}M_\lambda^3.
\end{eqnarray}
  This expression coincides exactly with Eq.~\eqref{eq:mass_eoms2}. However, Eq.~\eqref{eq:mass_eoms2}
is just a truncated Taylor expansion of the full order three EOMS result (Eq.~\eqref{eq:mass_eoms1}).
The EOMS result, automatically includes higher orders from the loop calculations, like the logarithmic and the arccosinus terms of Eq. \eqref{eq:mass_eoms1}. These terms, apart from reflecting the proper analytic dependence coming from the loops, have proved to lead to a faster chiral convergence in many cases \cite{Fuchs:2003ir,Lehnhart:2004vi,Schindler:2006it,Schindler:2006ha,Geng:2013xn,
MartinCamalich:2010fp,Geng:2008mf,Geng:2010df,Geng:2010yc}.

\section{Numerical results}
In Eq. \eqref{eq:mass_eoms1}, there are three low energy constants $c_1$, $c_7$ and $g_A$. As mentioned in Ref. \cite{Sun:2014aya}, $g_A$ can be fixed by comparing with other theoretical calculations. In Ref. \cite{Hu:2005gf}, the Lagrangian depicting doubly heavy baryon and meson interaction is constructed based on the heavy diquark symmetry
\begin{eqnarray}
\mathcal{L}=\text{Tr}[T^\dag_a(iD_0)_{ba}T_b]-g\text{Tr}[T_a^\dag T_b\vec{\sigma}\cdot \vec{A}_{ba}]+\cdots,
\end{eqnarray}
where $T_{a,i\beta}=\sqrt{2}(\Xi^*_{a,i\beta}+\frac{1}{\sqrt{3}}\Xi_{a,\gamma}\sigma^i_{\gamma \beta})$. By fitting the $D^{*+}$ decay width, one gets the coupling $g=0.6$. Comparing the Lagrangians with ours, we obtain $g_A=-g/3=-0.2$.

Besides the doubly heavy baryon mass $m$ in chiral limit, the low energy constants $c_1$ and $c_7$ still need to be determined. In order to do this, we fit the lattice data in Ref. \cite{Alexandrou:2012xk}. 

The masses of $\Xi_{cc}$ are given for different $m_\pi$ and $m_c$ in Ref. \cite{Alexandrou:2012xk}. The strange quark mass is tuned as to reproduced the kaon mass.  As in Ref.~\cite{Sun:2014aya}, we assume that only the bare mass $m$ depends on the 
valence charm quark mass $m_c$ and use the same ansatz as in Refs.~\cite{Alexandrou:2012xk,Sun:2014aya}
\begin{eqnarray}
m=\tilde{m}+2m_c+\alpha/m_c+\mathcal{O}(1/m_c^2).
\end{eqnarray}
The {\it physical} mass $m_c^{phy}$ is tuned to reproduce the mass of the $D$ meson at the physical point in Ref. \cite{Alexandrou:2012xk}. And here, for the different  lattice parameters $\beta=3.9$, $\beta=4.05$ and $\beta=4.2$, the values of physical charm quark mass $m_c^{phy}$ are given as $0.598\pm 0.066$, $0.591\pm 0.028$ and $0.598\pm 0.070$ GeV, respectively.
 The results of our fit to the lattice data are shown in Table. \ref{tab:best_fit_result}.

\begin{figure*}
\centering
\includegraphics[width=0.35\linewidth]{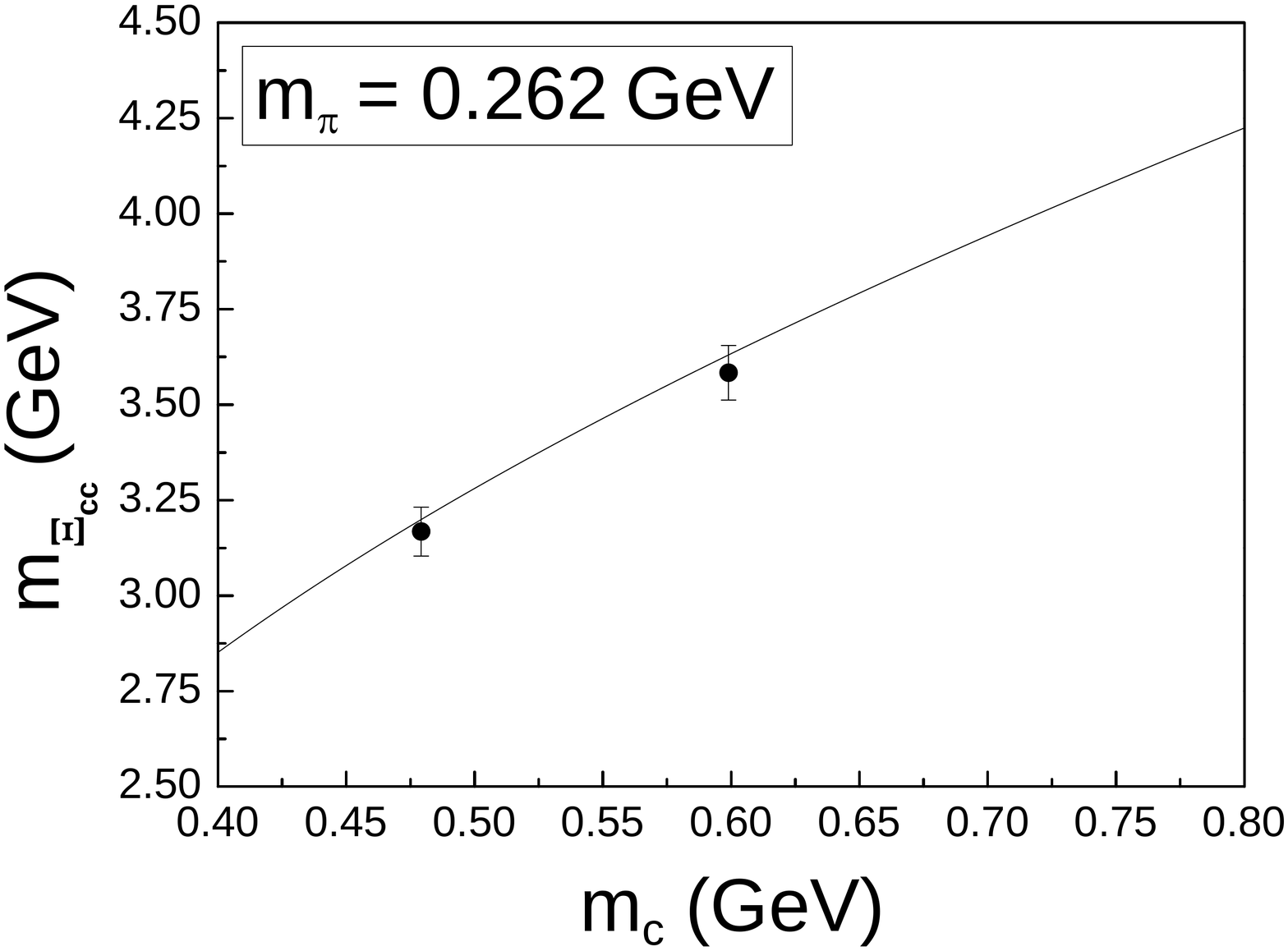}
\includegraphics[width=0.35\linewidth]{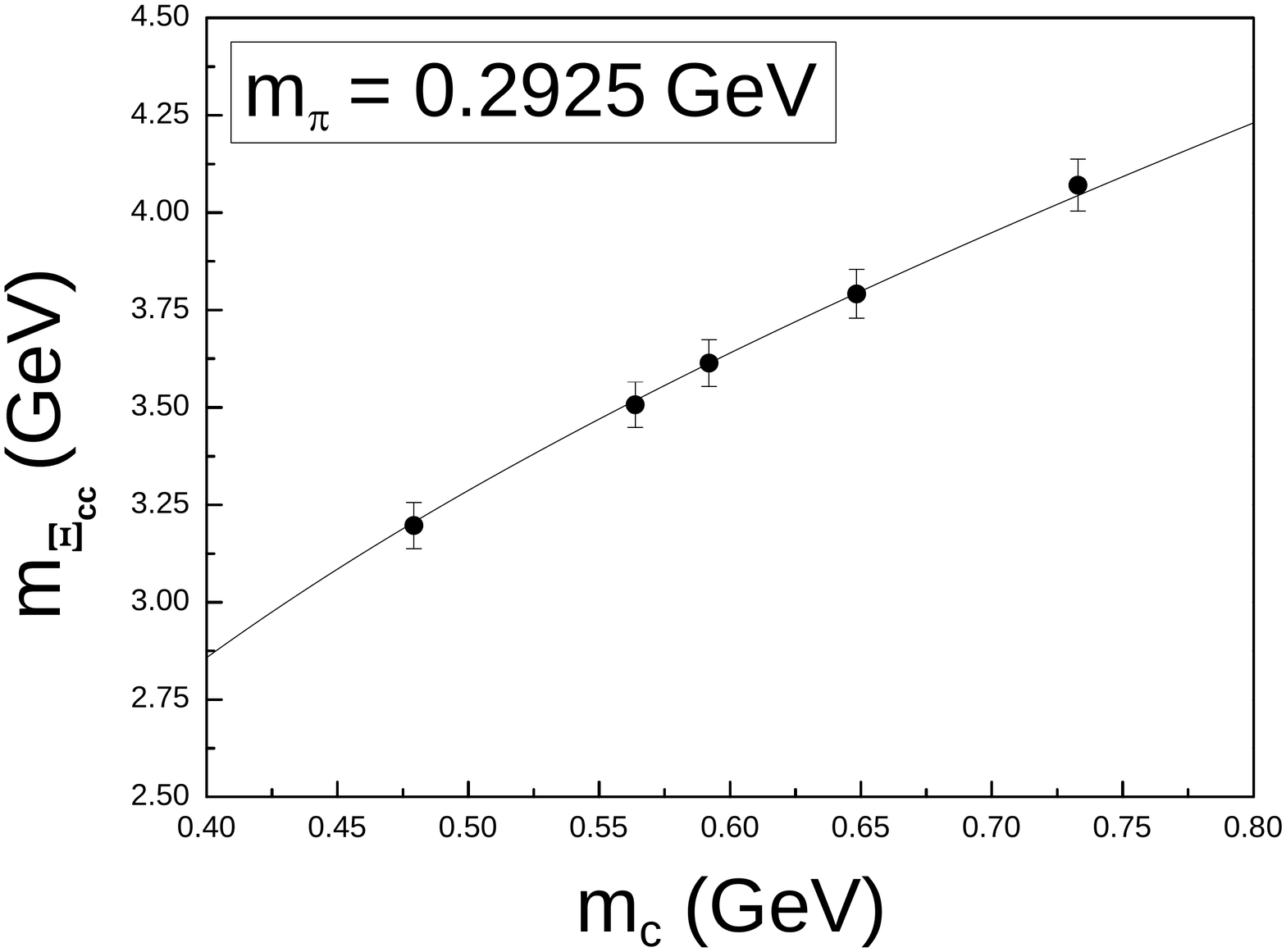}\\
\includegraphics[width=0.35\linewidth]{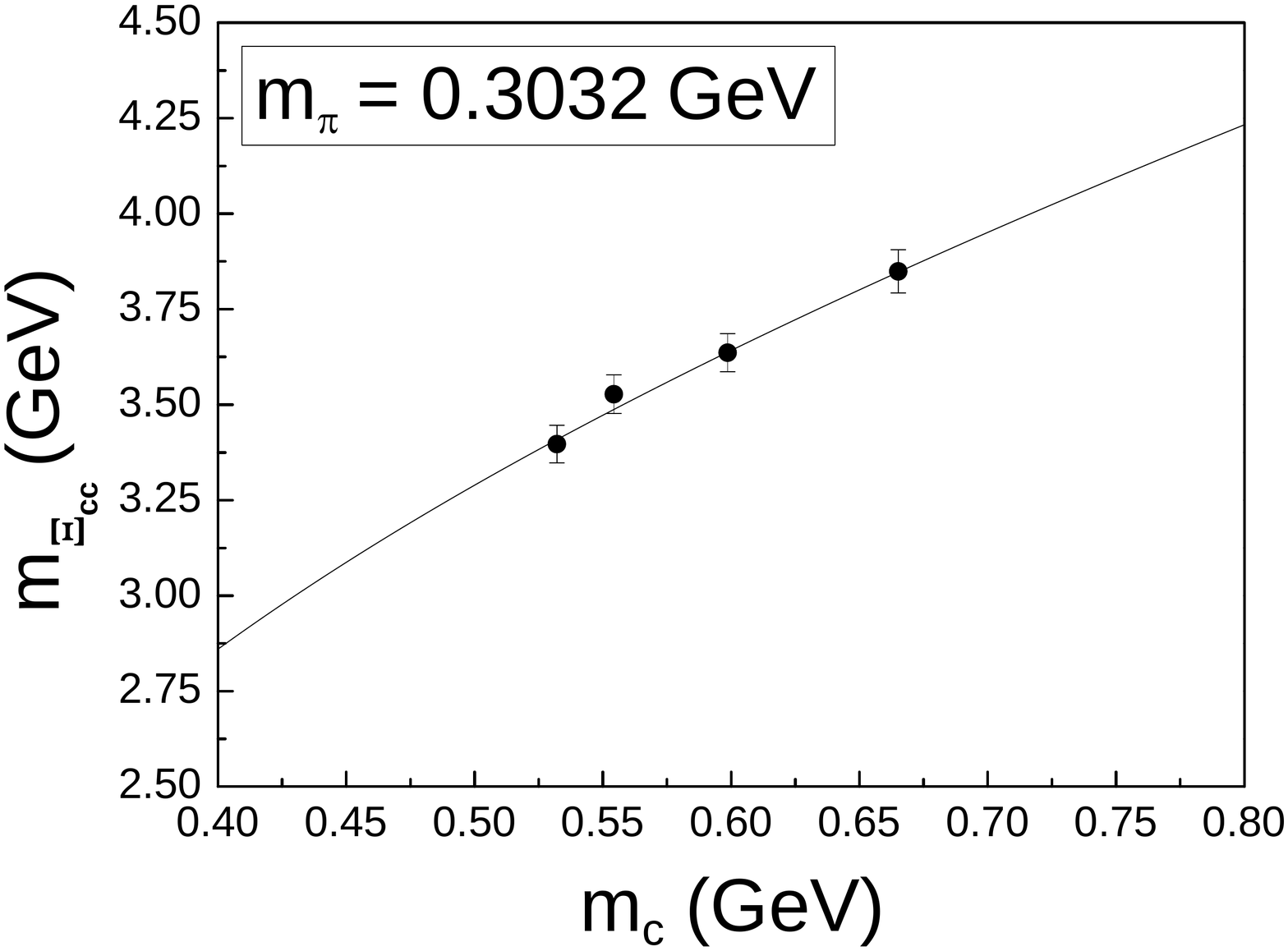}
\includegraphics[width=0.35\linewidth]{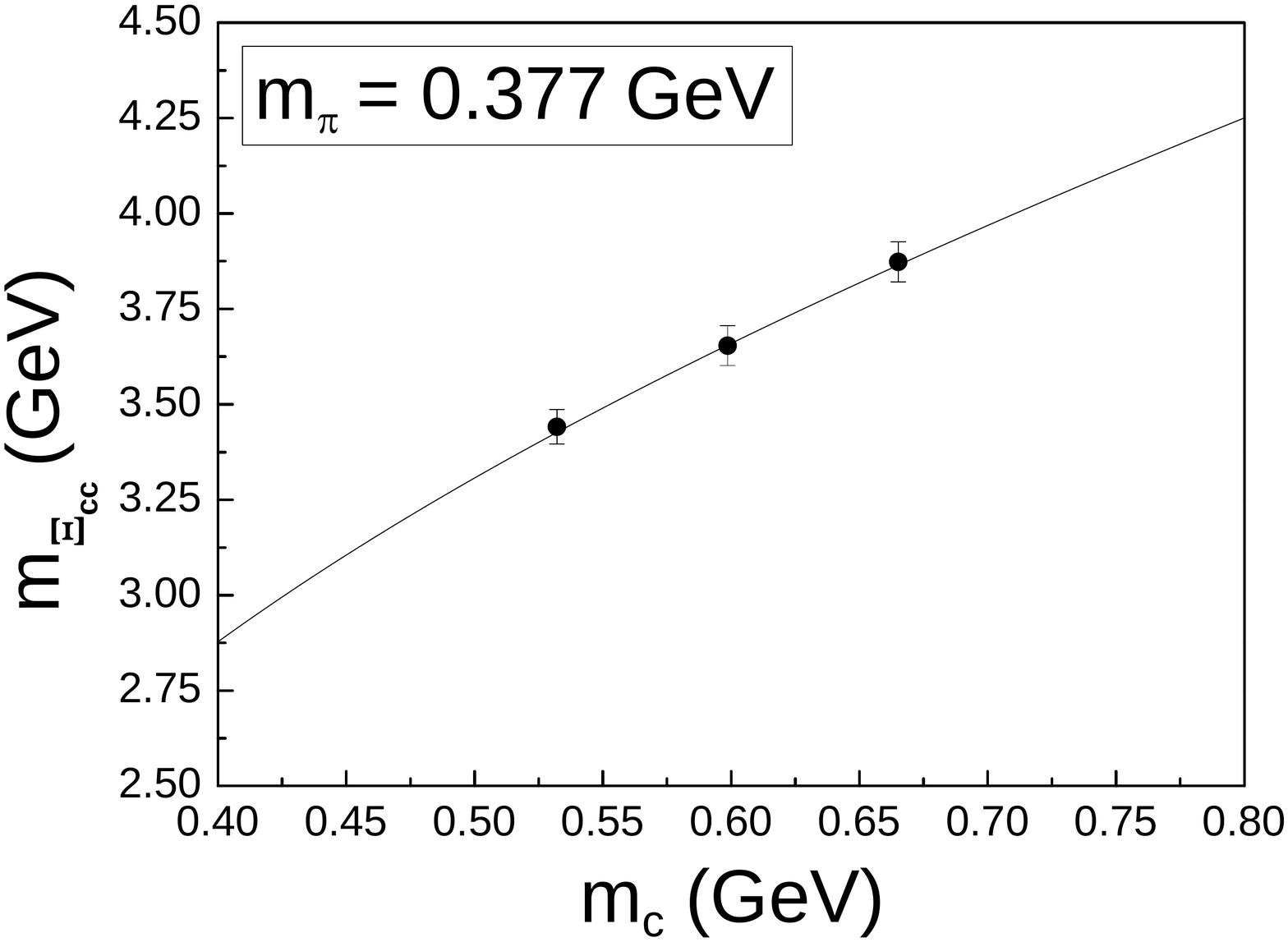}\\
\includegraphics[width=0.35\linewidth]{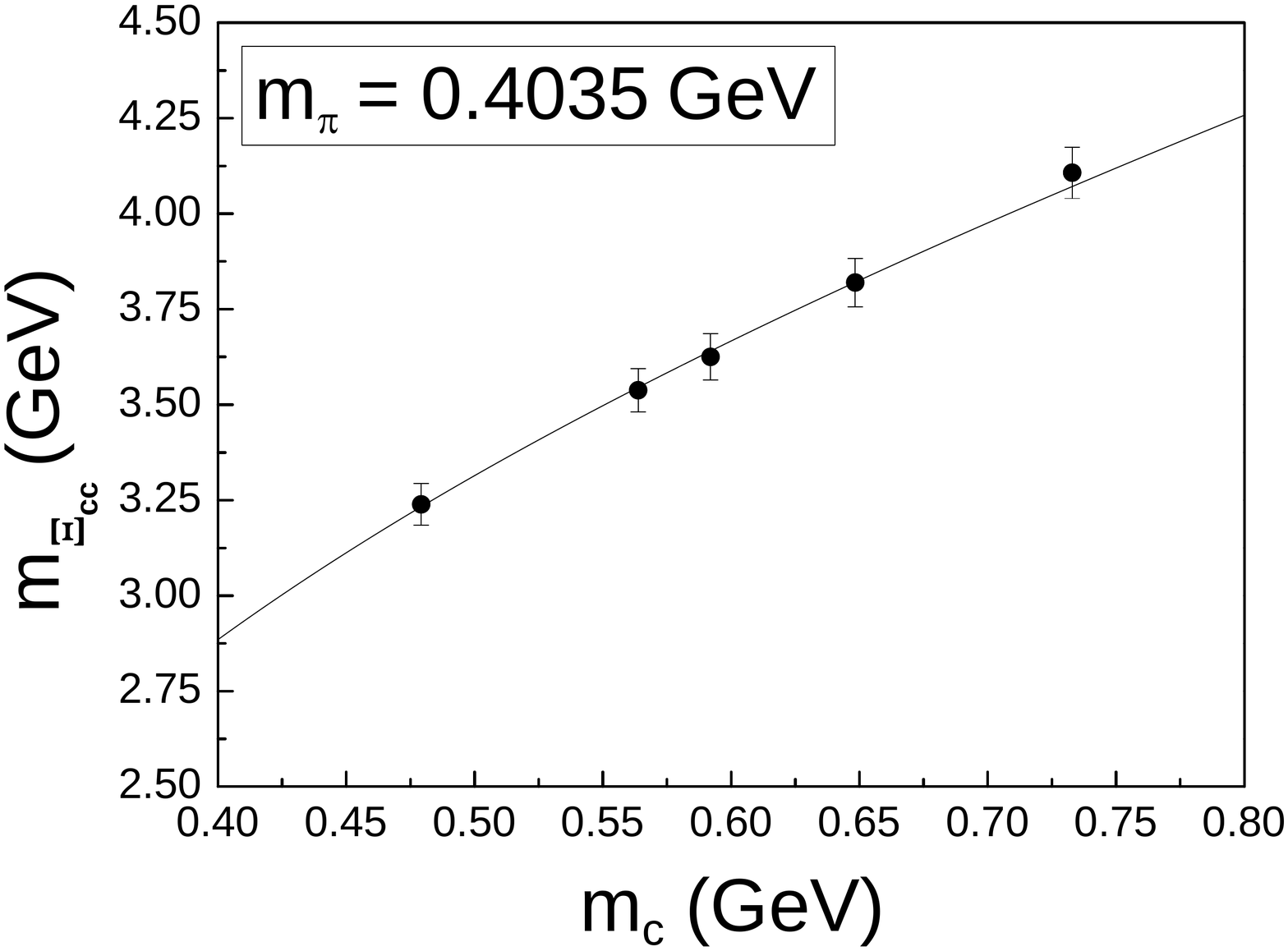}
\includegraphics[width=0.35\linewidth]{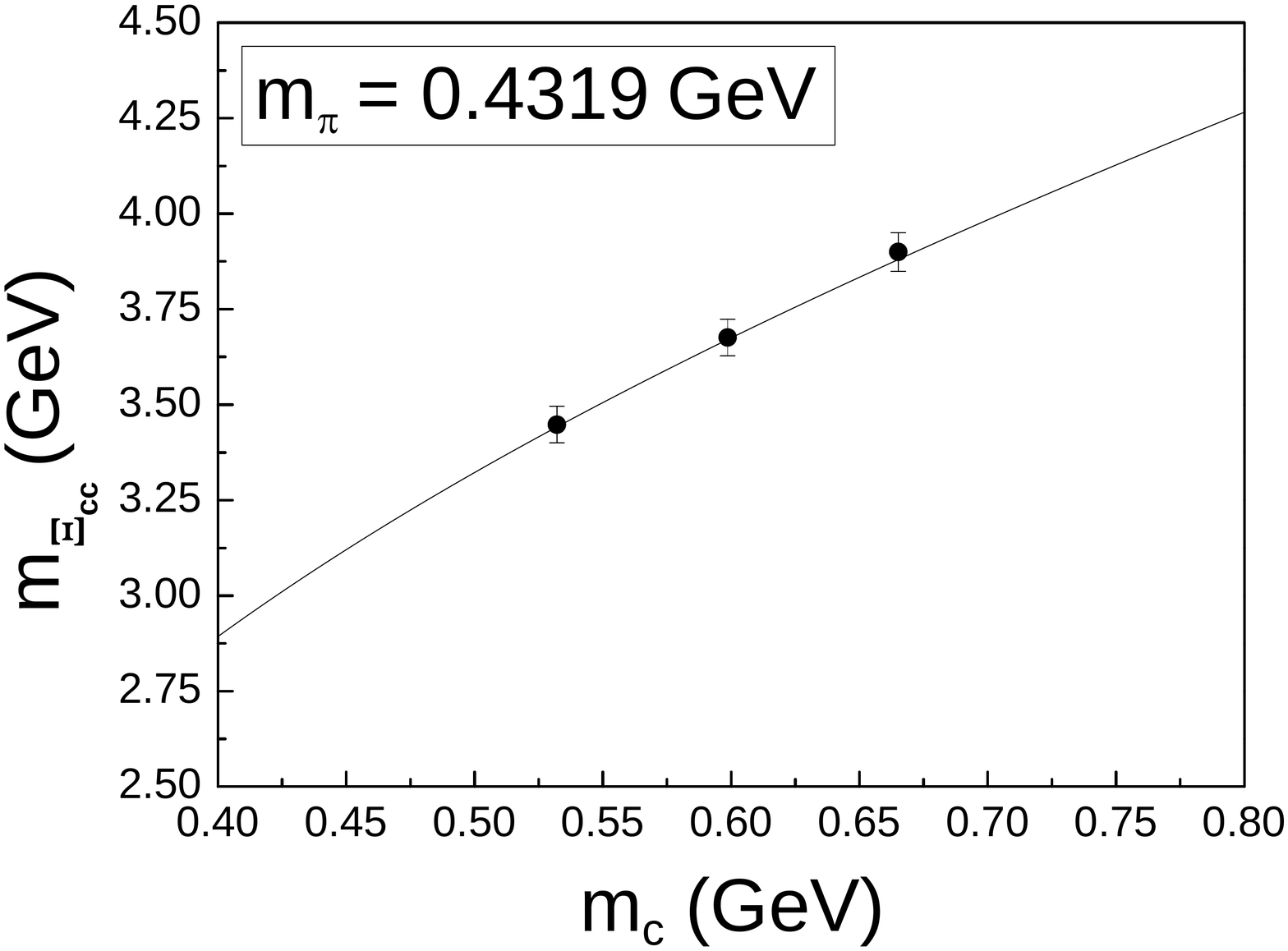}\\
\includegraphics[width=0.35\linewidth]{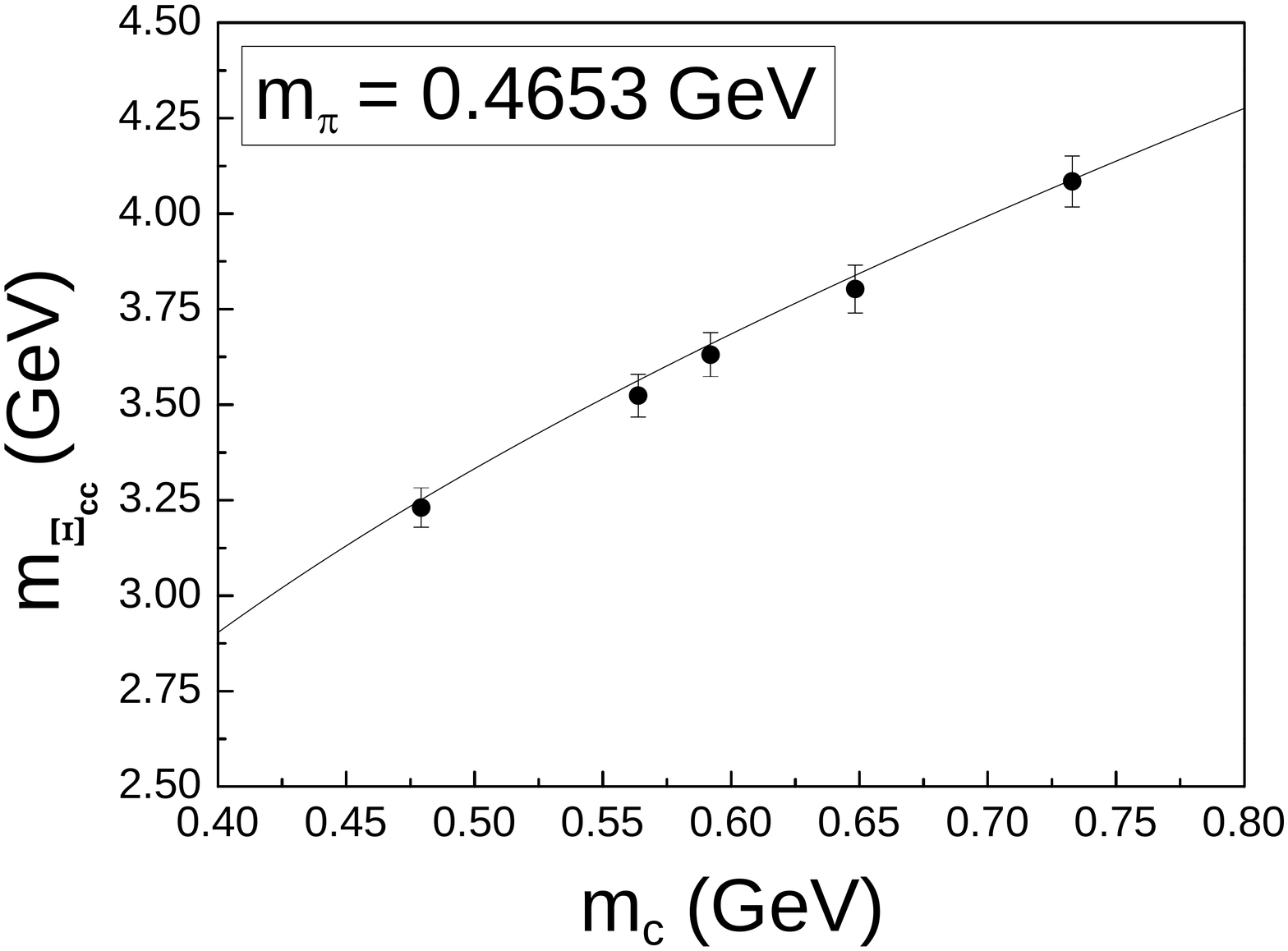}
\includegraphics[width=0.35\linewidth]{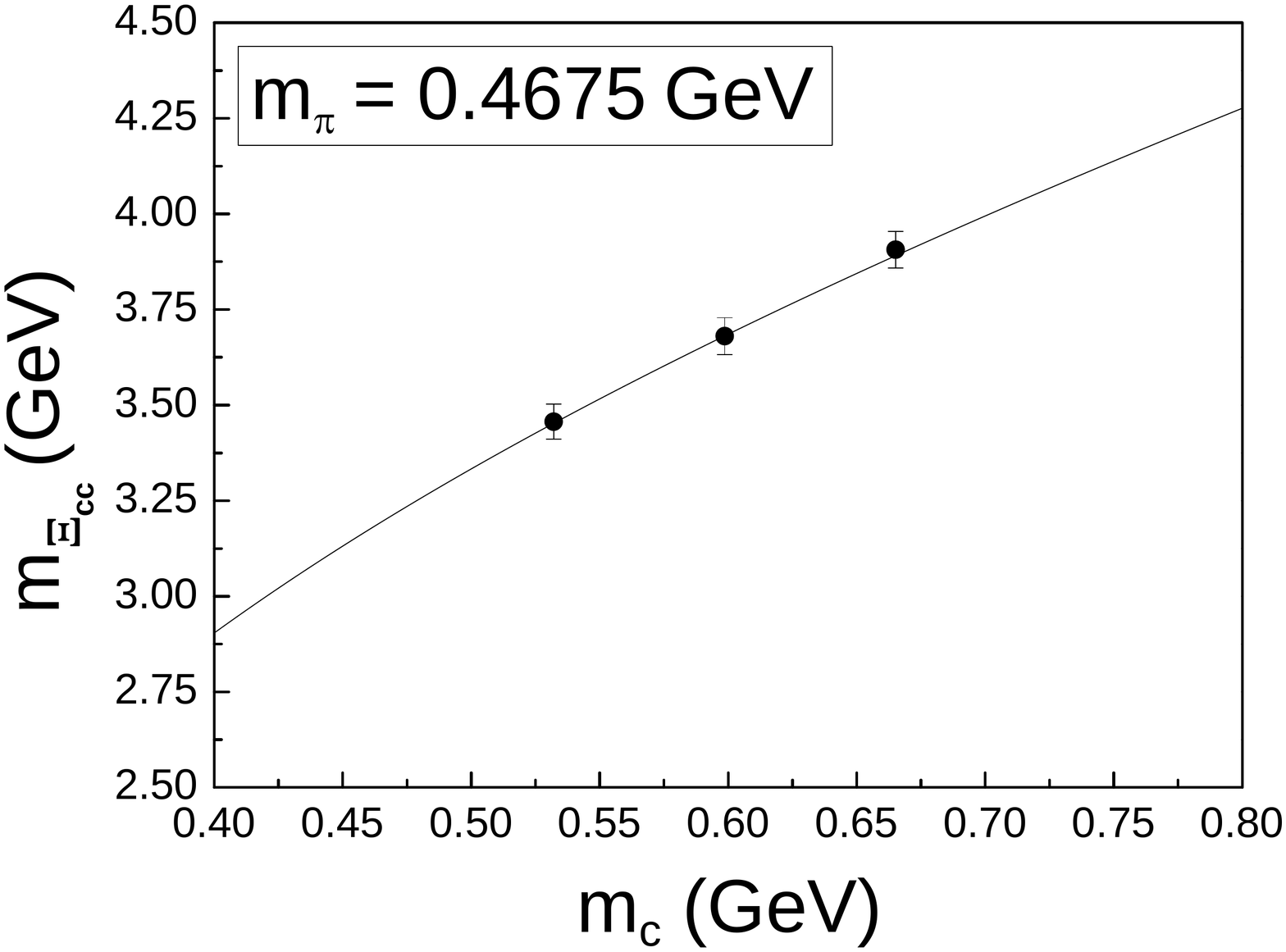}\\
\includegraphics[width=0.35\linewidth]{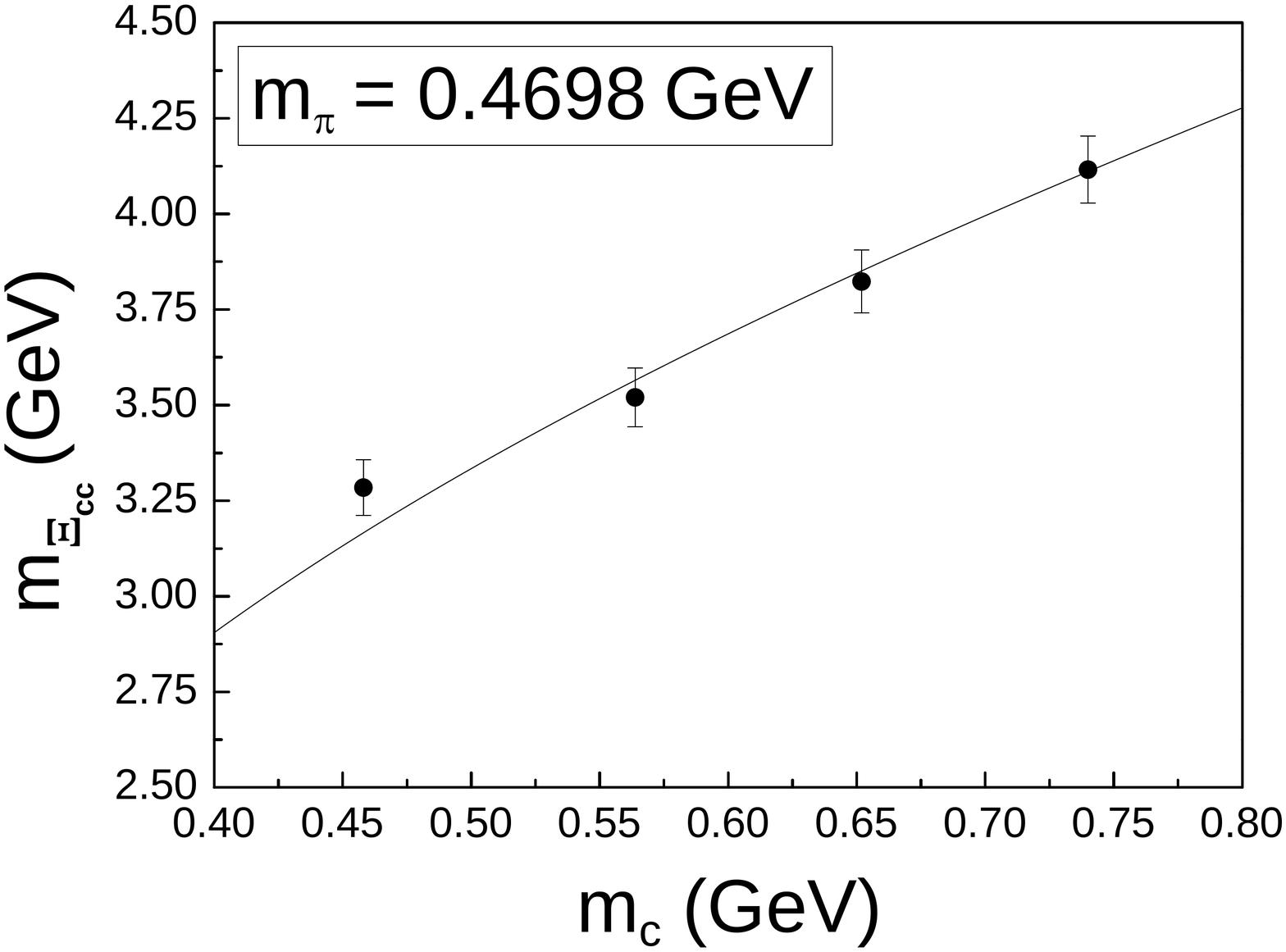}
\includegraphics[width=0.35\linewidth]{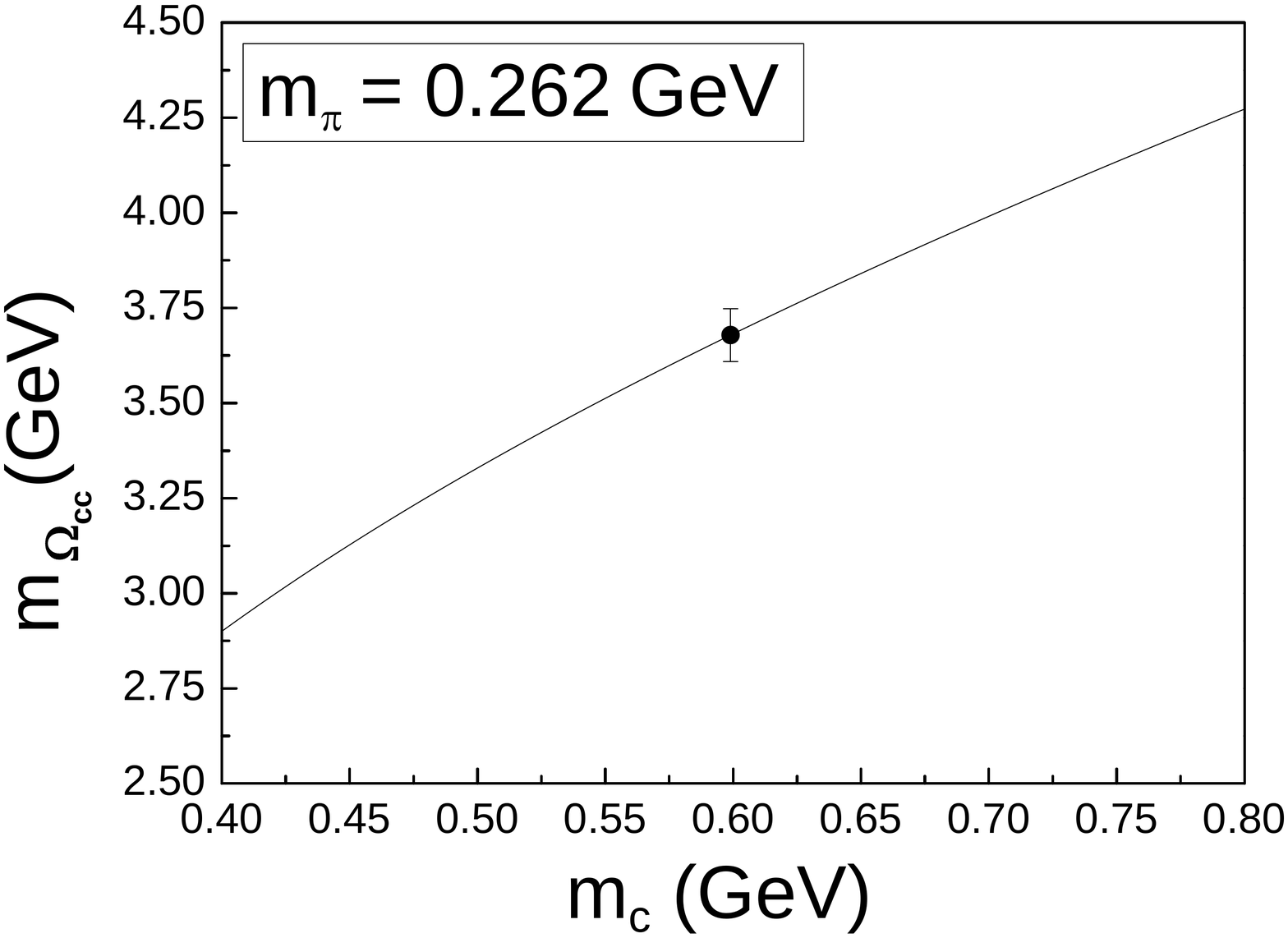}
\caption{Masses of $\Xi_{cc}$ and $\Omega_{cc}$ as a function of $m_c$ for different pion masses. The dots  are lattice data from Ref. \cite{Alexandrou:2012xk}, and the solid curves are our fitted result in the EOMS renormalization scheme. \label{fig:fit}}
\end{figure*}

In Fig. \ref{fig:fit}, we plot the best fit results.
% with $\tilde{m}=3.034$ GeV, $\alpha=-0.464$ GeV$^2$, $c_1=-0.267$ and $c_7=-0.185$
 The doubly heavy baryons masses are shown in Table. \ref{tab:doubly_heavy_baryon_mass} for the different values of the physical charm quark mass obtained by lattice QCD. Finally, the averages for these masses are $m_{\Xi_{cc}}=3.597\pm0.114$ MeV and  $m_{\Omega_{cc}}=3.652\pm0.118$ MeV.
%The uncertainties come from both the fit and that of charmed quark in lattice QCD calculation.

%\begin{eqnarray}
%m_{\Xi_{cc}}=3.593\pm 0.402\, \, \text{GeV},\,\, m_{\Omega_{cc}}=3.660 \,\,\text{GeV}.
%\end{eqnarray} 
\begin{table}
\caption{Values and uncertainties of the parameters $\tilde{m}$, $\alpha$, $c_1$ and $c_7$ obtained by fitting the lattice data from \cite{Alexandrou:2012xk}.\label{tab:best_fit_result}}
\begin{tabular}{c|ccccc}\toprule[1pt]
      &$\tilde{m}$ (GeV)&$\alpha$ (GeV$^2$)&$c_1$&$c_7$&$\chi^2_{d.o.f}$\\\midrule[0.5pt]
value &3.110&-0.459&-0.098&-0.073&\multirow{2}*{0.22}\\
error &0.111&0.047&0.045&0.089&\\\bottomrule[1pt]
\end{tabular}
\end{table} 

\begin{table}
\caption{Masses of $\Xi_{cc}$ and $\Omega_{cc}$ and their corresponding uncertainties for the different values of the charm quark mass from lattice QCD. All the values are expressed in GeV. \label{tab:doubly_heavy_baryon_mass}}
\begin{tabular}{c|c|c}\toprule[1pt]
 $m_c^{phy}$ &$m_{\Xi_{cc}}$&$m_{\Omega_{cc}}$ \\\midrule[0.5pt]
 0.598 $\pm$ 0.066& 3.608$\pm$ 0.218&3.663$\pm$0.223\\
 0.591 $\pm$ 0.028& 3.585$\pm$ 0.166&3.640$\pm$0.173\\
 0.598 $\pm$ 0.070& 3.608$\pm$ 0.225&3.663$\pm$0.230\\\bottomrule[1pt]
\end{tabular}
\end{table} 

 Our results, which are obtained by a chiral extrapolation from lattice data, although consistent with the SELEX measurement ($m_{\Xi_{cc}}=3519 \pm 1$ MeV) because of their large error bars, agree better with most theoretical estimates predicting a larger mass for this baryon. In Table VIII of Ref.~\cite{Karliner:2014gca}, a wide compilation of theoretical predictions can be found.

In Ref. \cite{Sun:2014aya}, using the heavy baryon approach, the same set of parameters is also determined by fitting the lattice data from \cite{Alexandrou:2012xk}. However, in this latter approach, one can not give any information of the low energy constant $c_1$, since the term corresponding to $c_1$ is the same constant for all cases and the term coming from the loop contribution does not depend on the baryon mass  (so that the term corresponding to $c_1$ can be absorbed into the baryon mass in the chiral limit).

\section{summary}
The doubly heavy baryons are very interesting hadronic systems although the experimental situation needs still to be settled. 
 The charmed quarks are relatively heavy so that they can  be treated as spectators. Consequently, the chiral dynamics is solely governed by the light quark.
In this work, we have used an effective Lagrangian which describes the chiral dynamics of doubly heavy baryons to calculate  chiral corrections to their masses. In order to deal with the power counting problem, intrinsic to baryon ChPT, we have used the EOMS method. Within this scheme, we have obtained the baryon masses up to N$^2$LO. We have shown that a truncation of our results reproduces those of the heavy baryon approach.  

We have also performed a numerical analysis of the doubly heavy baryon masses. From  the $D^{*+}$ decay width one gets the coupling $g_A=- 0.2$. Then, by fitting the lattice data, at several pion and charm quark masses, from Ref. \cite{Alexandrou:2012xk},  the parameters  $c_1$, $c_7$, $\tilde{m}$ and $\alpha$ have been determined. Consequently, the masses of the doubly charmed baryons at the physical point have been predicted. The use of the EOMS scheme  has allowed us to determine the constant $c_1$ which in the heavy baryon method is fully correlated with the baryon mass in the chiral limit and cannot be disentangled. We expect that, as in many other observables, the chiral convergence of EOMS will be faster than in the heavy baryon  approach. 
  We are looking forward to more developments of experimental and theoretical studies in this field, through which we can deepen our understanding of the hadron spectrum and nonperturbative QCD.

\section*{Acknowledgments}
We would like to thank Astrid Hiller Blin and En Wang for their help and useful discussions.
This research has been supported by
the Spanish Ministerio de Econom\'ia y Competitividad (MINECO)  and European FEDER funds under Contracts No. FIS2011-28853-C02-01, FIS2014-51948-C2-2-P and SEV-2014-0398 and by Generalitat Valenciana under Contract PROMETEOII/2014/0068.

\section*{Appendix: Loop integrals}

In this appendix, we give the loop integrals which are needed in our calculation.

\begin{eqnarray}
I_a(0) &=& i\int \frac{d^nk}{(2\pi^n)}\frac{1}{k^2-m^2+i\epsilon}=2m^2\bar{\lambda},\nonumber
\end{eqnarray}
\begin{eqnarray}
I_\lambda(0) &=& i\int \frac{d^nk}{(2\pi^n)}\frac{1}{k^2-M_\lambda^2+i\epsilon}\nonumber\\
&=&2M_\lambda^2\bar{\lambda}+\frac{M_\lambda^2}{(4\pi)^2}\ln\frac{M_\lambda^2}{m^2},\nonumber
\end{eqnarray}
\begin{eqnarray}
I_{\lambda a}(-p,0)&=&i\int \frac{d^nk}{(2\pi^n)}\frac{1}{(p-k)^2-m^2+i\epsilon}\frac{1}{k^2-M_\lambda^2+i\epsilon}\nonumber\\
&=&2\bar{\lambda}+\frac{1}{(4\pi)^2}\left(-1+\frac{p^2-m^2+M_\lambda^2}{2p^2}\ln \frac{M_\lambda^2}{m^2}+f_0\right),\nonumber
\end{eqnarray}
where 

\begin{eqnarray}
\bar{\lambda}&=&\frac{1}{(4\pi)^2}\left[-\frac{1}{\epsilon}-\frac{1}{2}\left(\ln \frac{4\pi}{m^2}+1-\gamma\right)\right],\nonumber\\
f_0&=&\left\{
\begin{array}{c}
\frac{\sqrt{\Theta^2-\Delta^2}}{p^2}\arccos \frac{-\Delta}{\Theta},\,-1<\frac{\Delta}{\Theta}<1\\ 
\frac{\sqrt{\Delta^2-\Theta^2}}{2p^2}\ln\frac{\Delta+\sqrt{\Delta^2-\Theta^2}}{\Delta-\sqrt{\Delta^2-\Theta^2}}, \, \frac{\Delta}{\Theta}<-1\\ 
\frac{\sqrt{\Delta^2-\Theta^2}}{2p^2}\ln\frac{\Delta+\sqrt{\Delta^2-\Theta^2}}{\Delta-\sqrt{\Delta^2-\Theta^2}}-i\pi\frac{\sqrt{\Delta^2-\Theta^2}}{p^2}, \, \frac{\Delta}{\Theta}>1 \\
0,\, \frac{\Delta}{\Theta}=\pm 1
\end{array} \right.\nonumber
\end{eqnarray}
with $\Delta=p^2-m^2-M_\lambda^2$ and $\Theta=2mM_\lambda$.

\bibliographystyle{plain}

\end{document}